
\magnification=\magstep1
\headline={\ifnum\pageno=1\hfil\else\hfil\tenrm--\ \folio\ --\hfil\fi}
\footline={\hfil}
\hsize=6.0truein
\vsize=8.54truein
\hoffset=0.25truein
\voffset=0.25truein
\baselineskip=20pt
%
%
\tolerance=9000
\hyphenpenalty=10000
%
%
%

\font\mbf=cmmib10 \font\mbfs=cmmib10 scaled 833
\font\msybf=cmbsy10 \font\msybfs=cmbsy10 scaled 833

%
%
%

\textfont9=\mbf \scriptfont9=\mbfs \scriptscriptfont9=\mbfs

\textfont10=\msybf \scriptfont10=\msybfs \scriptscriptfont10=\msybfs
%
%
\mathchardef\alpha="710B
\mathchardef\beta="710C
\mathchardef\gamma="710D
\mathchardef\delta="710E
\mathchardef\epsilon="710F
\mathchardef\zeta="7110
\mathchardef\eta="7111
\mathchardef\theta="7112
\mathchardef\iota="7113
\mathchardef\kappa="7114
\mathchardef\lambda="7115
\mathchardef\mu="7116
\mathchardef\nu="7117
\mathchardef\xi="7118
\mathchardef\pi="7119
\mathchardef\rho="711A
\mathchardef\sigma="711B
\mathchardef\tau="711C
\mathchardef\upsilon="711D
\mathchardef\phi="711E
\mathchardef\chi="711F
\mathchardef\psi="7120
\mathchardef\omega="7121
\mathchardef\varepsilon="7122
\mathchardef\vartheta="7123
\mathchardef\varpi="7124
\mathchardef\varrho="7125
\mathchardef\varsigma="7126
\mathchardef\varphi="7127
\mathchardef\nabla="7272
\mathchardef\cdot="7201
%
\def\spose#1{\hbox to 0pt{#1\hss}}
\def\lta{\mathrel{\spose{\lower 3pt\hbox{$\mathchar"218$}}
     \raise 2.0pt\hbox{$\mathchar"13C$}}}
\def\gta{\mathrel{\spose{\lower 3pt\hbox{$\mathchar"218$}}
     \raise 2.0pt\hbox{$\mathchar"13E$}}}
%
%

%
%

%
%

%
%
%

\def\cm{{\rm\,cm}}

\def\kms{{\rm\,km\,s^{-1}}}

\def\etal{et al.$~$}

\def\Sec{\S $~$}

%
%

%
%
\newcount\eqnumber
\eqnumber=1
%
\def\new{{\the\eqnumber}\global\advance\eqnumber by 1}
%
%
\def\ref#1{\advance\eqnumber by -#1 \the\eqnumber
     \advance\eqnumber by #1 }
%
%
\def\last{\advance\eqnumber by -1 {\the\eqnumber}\advance
     \eqnumber by 1}
%
%
\def\eqnam#1{\xdef#1{\the\eqnumber}}
%
%
%
\def\refindent{\par\noindent\hangindent=3pc\hangafter=1 }
\def\aa#1#2#3{\refindent#1, A\&A, #2, #3}

\def\apj#1#2#3{\refindent#1, ApJ, #2, #3}
\def\apjsup#1#2#3{\refindent#1, ApJS, #2, #3}

\def\mnras#1#2#3{\refindent#1, MNRAS, #2, #3}

\def\nature#1#2#3{\refindent#1, Nature, #2, #3}

\def\refbook#1{\refindent#1}

%
%
\def\refrule{\hbox to 3pc{\leaders\hrule depth-2pt height 2.4pt\hfill}}
%
%
%
\def\sect#1 {
  \vskip 1. truecm plus .2cm
  \bigbreak
  \centerline{\bf #1}
  \nobreak
  \bigskip
  \nobreak}
\def\subsec#1#2 {
  \bigbreak
  \centerline{#1.~{\bf #2}}
  \bigskip}
%
%

\def\lya{{\rm Ly}\alpha}
\def\fnhi{f(N_{\rm HI})}
\def\nhi{N_{\rm HI}}

\def\ocdm{\Omega _{\rm CDM}}
\def\ohdm{\Omega _{\rm HDM}}
\def\olambda{\Omega _{\Lambda}}
\def\obaryon{\Omega _{\rm b}}
\def\rv{r_{\rm vir}}
\def\gtsima{$\; \buildrel > \over \sim \;$}
\def\ltsima{$\; \buildrel < \over \sim \;$}
\def\prosima{$\; \buildrel \propto \over \sim \;$}
\def\gsim{\lower.5ex\hbox{\gtsima}}
\def\lsim{\lower.5ex\hbox{\ltsima}}
\def\simgt{\lower.5ex\hbox{\gtsima}}
\def\simlt{\lower.5ex\hbox{\ltsima}}
\def\simpr{\lower.5ex\hbox{\prosima}}
\def\rd {{\rm d}}

\parskip .15cm plus .1cm
\null\vskip 1.cm

\centerline{\bf DAMPED LYMAN ALPHA SYSTEMS AND GALAXY FORMATION}

\bigskip
\centerline{ H. J. Mo$^1$ and J. Miralda-Escud\'e$^{1,2}$}
\medskip
\centerline{$^ 1$ Institute of Astronomy}
\centerline{Madingley Road, Cambridge, CB3 0HA, UK}
\medskip
\centerline{$^ 2$ Institute for Advanced Study}
\centerline{ Olden Lane, Princeton, NJ 08540}

\centerline{submitted to ApJL, {\it Received:} 1994, .........} \par

\vskip 1.cm

\sect{ABSTRACT}

  We examine the constraints on theories of galaxy formation that are
obtained from observations of damped $\lya$ (DL) systems, assuming they
are gaseous protodisks in dark matter halos. Using the Press-Schechter
formalism, we find that the mixed dark matter model, with $\ohdm =
0.3$, $\ocdm = 0.65$, $\obaryon = 0.05$, and $h=0.5$, is ruled out
because the number of galactic halos at $z\simeq 3$ is too small to
account for the total gaseous mass in DL systems, even under the
assumption that all the gas in collapsed halos has settled into disks
of neutral gas. The standard CDM model can account for the gas in DL
systems if the bias is $b\lsim 2$; the same is true for the CDM model
with a cosmological constant, if $b\lsim 1.5$ for $\Lambda = 0.8$.
However, one still needs to assume that a fraction $\gsim 0.4$ of the baryons
in collapsed halos at $z\simeq 3$ is in the form of neutral gas in
disks. We also calculate the column density distribution $f(\nhi)$ of
the DL systems, in terms of the surface density profiles of disks and
the distribution of their central column densities. It is shown that the
form of $f(\nhi)$ at the high end of column density is a diagnostic for
the nature of DL systems.

\noindent{{\it Subject headings}: galaxies: formation - QSO: absorption
lines - cosmology}

\vfill\eject

  \sect {1.~ INTRODUCTION}

  Damped $\lya$ absorption systems are, at present, the best
observational probe we have to study galaxy formation and evolution at
high redshift (see Wolfe 1993). The observed systems are selected only
because they lie on the line-of-sight to an unrelated quasar, rather
than from any special property such as a high optical or radio
luminosity. Thus, damped systems are probably the progenitors of typical
galaxies. In this paper, we show that the observations of DL systems can
be used as a constraint for models of galaxy formation. In \Sec 2 we
discuss how the column density distribution of DL systems can be used to
infer the typical central column densities of the objects producing
them. In \Sec 3 we investigate the number of DL systems and the total
gaseous mass they contain, as predicted by currently favored models of
galaxy formation. The results are discussed in \Sec 4.

  \sect {2.~ THE COLUMN DENSITY DISTRIBUTION FOR DISKS}

  In this section, we calculate the column density distribution of DL
systems that results from thin disks of HI gas. We write the surface HI
density profile of a disk as
$$
\Sigma (r)=\Sigma _0 F(r/r_0) ,
\eqno(1)
$$
where $r$ is the radius on the disk, $\Sigma _0$ is the central surface
density, and $r_0$ is a typical scalelength; the function $F(r/r_0)$ is
a decreasing function of $r$, and is normalized to $F(0)=1$. If the
normal to a disk is inclined to an angle $\theta$ with respect to the
line of sight, the cross section for intersecting the disk at radius $r$
is $2\pi r\, dr\, \cos(\theta)$, and the column density observed is
$\nhi=N_0\, F(r/r_0)/\cos(\theta)$. For randomly oriented disks, $\cos
(\theta)$ is uniformly distributed, and the total cross section for
intersecting a disk with column density larger than $\nhi$ is
$$
\sigma (\nhi) =2\pi \int _0^\infty r\,{\rm d}r \int _0^a
\cos \theta\, {\rm d}\cos\theta ,\eqno(2)
$$
where $a={\rm min}\left\lbrack 1, N_0\, F(r/r_0)/\nhi\right\rbrack $.
Differentiating with
respect to $\nhi$, we obtain:

$$
{{\rm d}\sigma \over {\rm d}\nhi}= {2\pi\, r_0^2\, N_0^2 \over \nhi ^3}
\int _{x_1}^\infty x{\rm d}x [F(x)]^2 ,
\eqno(3) $$
where $x_1$ is given by $F(x_1)={\rm min}[1,\nhi/N_0]$.

If the number density of systems of central column density between $N_0$
and $N_0 + {\rm d}N_0$ is
$n (N_0, z)\, {\rm d}N_0$, then the number of absorption
lines found per unit redshift and per unit column density is
$$
f(\nhi, z)={{\rm d}l\over {\rm d}z} \int {{\rm d}\sigma \over
{\rm d}\nhi} n (N_0, z){\rm d}N_0 , \eqno(4) $$
where $l(z)$ is the comoving distance at redshift $z$ (see Sargent
\etal 1980). If the systems have a distribution of scalelengths, then
one simply needs to substitute $r_0^2$ in equations (3) by its average
for all systems of a fixed central column density. When the column
density $\nhi$ is higher than $N_0$ for all the disks, then the HI
column density distribution will be
$$
f(\nhi) \propto \left({N_0 \over \nhi}\right) ^3
\eqno(5)
$$
(see Barcons \& Fabian 1987) independently of the forms of $F$ and $n$.
For lower column densities, the $\nhi$ distribution depends on the
profiles $F(r/r_0)$ and the distribution of $N_0$. For exponential disks
with $F(r/r_0) = {\rm exp}(-r/r_0)$, and $N_0 = {\rm const.}$, the HI column
density distribution for $\nhi < N_0$ is
$$
f(\nhi) \propto {N_0\over \nhi}
\left\lbrack 1+{\rm ln} \left({N_0 \over \nhi }\right) ^2\right\rbrack .
\eqno(6)
$$
In Figure 1, we show the observed HI column density distribution,
$\fnhi$ taken from Lanzetta \etal (1991) and Tytler (1987). The curves
are the distribution function derived for exponential disks from
equations (3), assuming that $N_0$ is the same for all disks, for the
cases $N_0 = 10^{21},\,10^{21.5},\,10^{22} {\rm cm}^{-2}$.
The amplitudes of the curves are fitted to the observed
points.

  At present, the central surface brightness
of normal galaxies are remarkably constant, with the value $\sim 140
L_\odot {\rm pc}^{-2}$ (Freeman 1970), or a column density
$\sim 10^{23}{\rm proton}\,\,\cm^{-2}$ assuming a mass-to-light
ratio $\Upsilon \sim 5\Upsilon _\odot$ for disks. If the
DL systems have similar surface densities {\it in gas}, no steepening of
the distribution of $\nhi$ should have been observed yet. But if most
DL systems had much lower central column densities, a turnover of
$f(\nhi)$ could be present in the observed range. In fact, if the
gaseous disks at high redshift had profiles similar to the present HI
disks, a turnover should already have been observed (see Rao \& Briggs
1993). The point in Figure 1 at the highest column density may be an
indication of such a turnover, and therefore of a large number of
disks with $N_0 \simeq 10^{21.5}$ at high redshift. The statistics,
however, are not yet conclusive.

\sect {3.~ CONSTRAINING MODELS BY THE OBSERVATION OF DL SYSTEMS}

  According to most current models, the formation of structure in the
universe occurred through the growth of small inhomogeneities via
gravitational instability. In this scenario, galaxies are assumed to
form by cooling and condensation of gas in the potential wells of dark
halos (e.g., White \& Rees 1978). In this section we use the
Press-Schechter formalism (Press \& Schechter 1974) to estimate the
abundance of dark halos in different models. We then combine this with
some plausible assumptions on galaxy formation to predict the observed
number of DL systems and the gaseous mass they contain.

  We consider three different models for the formation of structure.
Model 1 is the standard CDM model, with $\ocdm = 0.95$, $\obaryon =0.05$
and $h=0.5$ (here and in the following we adopt a baryon density
parameter $\obaryon=0.0125\, h^{-2}$ from primordial nucleosynthesis,
where $h$ is the Hubble constant $H_0$ divided by $100\kms {\rm
Mpc}^{-1}$, see Walker \etal 1991). Model 2 has a cosmological constant
$\olambda = 0.78$, $\ocdm=0.2$, $\obaryon =0.02$ and $h=0.8$; we call
this model CDM$\Lambda$. In both of these models, we use the fit to the
CDM power spectrum in equation (G3) of Bardeen \etal (1986). Model 3
(called MDM; mixed dark matter) has $\ocdm=0.65$, $\ohdm=0.3$, $\obaryon
=0.05$ and $h=0.5$. For this model we use the power spectrum in equation
(1) of Klypin et al. (1993). The biasing parameters are varied for all
models, and are chosen to bracket values currently favored by
observations. These three models are of special interest in the
``post-COBE" epoch of cosmogony (see Bond 1993 for a review).

  In the Press-Schechter formalism, the comoving number density of halos
as a function of halo mass $M$ and redshift $z$ can be written as
$$
n(M,z)\, {\rm d} M={-3\over (2\pi)^{3/2}} {1\over r_0^3}
{\delta _c\over \Delta^2(r_0,z)}
{\rm exp} \left\lbrack -{\delta _c^2\over 2\Delta (r_0,z)}\right\rbrack
{\rd \Delta (r_0,z)\over \rd M}\rd M ~,
\eqno(7)
$$
where $M$ is the mass in a top-hat window with radius $r_0$
(comoving radius in present unit): $M=4\pi \rho_0 r_0^3/3$,
with $\rho_0$ being the present cosmic mass density in matter.
The quantity $\Delta (r_0,z)$ is the rms mass fluctuation
in a top-hat window with comoving radius $r_0$, linearly evolved
to redshift $z$, and is determined by the inital power spectrum.
The threshold $\delta _c$ is chosen to be 1.68 irrespective
of models (see White \etal 1993 for a discussion).
Assuming the spherical collapse model, one can
define a virialized radius, $\rv$, for each halo virialized at $z$:
$$
\rv=r_0(1+z)^{-1}(1+178\Omega_0^{-0.6})^{-1/3} ~,
\eqno(8)
$$
where $\Omega_0$ is the present-day density parameter in matter. This
radius is assumed to define the region within which the total virialized
mass of an object should be calculated (White \etal 1993). The circular
velocity $v_c$ for a virialized halo can then be written as:
$$
v_c\equiv \left( {G M\over \rv}\right)
^{1/2}=2^{-1/2}H_0r_0\Omega_0^{1/2}(1+z)^{1/2}
\left\lbrack 1+178 \Omega _0^{-0.6}\right\rbrack ^{1/6}.
\eqno(9)
$$
We shall calculate the abundance of virialized halos as a function of
circular velocity and redshift, using equations (7, 9).

Figure 2 shows the density of baryons
(scaled to zero redshift, in unit of present critical density)
in virialized
halos as a function of redshift. This is calculated from equation
(7) by assuming that the fraction of baryons in dark halos
is the same as the global value in each model. The results
are shown for three different intervals of halo circular velocities, as
shown in the figures.
If DL systems are due to neutral gas in systems which
form normal spiral galaxies, then
the relevant interval may be $v_c=100-250\kms$.
Gaseous disks might also form in systems of lower circular velocity, but
probably not below $v_c=50\kms$, since smaller objects would have a long
cooling time due to the ionizing background (Efstathiou 1992).
The upper limit corresponds to the maximum observed circular velocities
in spiral galaxies, $\sim 250\kms$, but this is also uncertain since
gaseous disks might survive
for some time after their host halo merges with a more massive one.
Clearly, the curves should be interpreted only as upper limits to the
mass that may be present in DL systems, since some of the gas in halos
may be ionized, or form stars rapidly, or be expelled from the halos in
winds from supernova explosions. The data points
shown by solid circles are the observed cosmic mass density
of HI gas in DL systems, $\Omega_{\rm D}$, adopted from Lanzetta \etal
(1993) for the cosmological models in consideration.
The figure shows that the baryonic mass in virialized
halos at high redshifts depends strongly on the bias
parameter. The large excess of the predicted $\Omega _{\rm D}$ at
low redshifts is not a problem for the models, since most of the gas has
probably been used up to form stars at present. In fact, observations of
present galaxies suggest that the mass of neutral gas in galaxies is
only $\sim 10\%$ of the total mass in stars.

  At high redshifts, the baryonic mass in galactic halos may still be
larger than that observed in DL systems in both the CDM and CDM$\Lambda$
models, with the bias parameter given in the figures. For the MDM model,
however, the baryonic mass in virialized halos is not enough to explain
the DL systems. The deficit is very large for the model with $b=1.5$, a
case favored by a normalization according to COBE results (see e.g.,
Klypin \etal 1993; Jing \etal 1993). So if DL systems are indeed due to
gas in galactic-sized virialized halos, then the MDM model based on the
COBE normalization can be ruled out.

  Similar conclusions have been reached recently by Subramanian \&
Padmanabhan (1994), although their values for $\Omega_{\rm D}$
predicted by the
models are lower than ours, because they assume a larger mass for the
halos giving rise to damped systems.

  Figure 3 shows the comparison of the predicted rate of incidence with
that observed. The data points are adopted from Lanzetta \etal (1993).
The number of absorption systems above a given column density depends
not only on the mass of gas in each halo, but also on its spatial
distribution, and is consequently much more uncertain and model
dependent than the total mass in damped systems. Here, we will assume
that the gaseous disks causing the DL systems have exponential profiles
with scalelengths that are the same as those of stellar disks in spiral
galaxies at the present time, for a fixed circular velocity. Assuming
also the Tully-Fisher relationship, and a constant central surface
brightness for spiral disks, the scalelength is $R_0 = 3.5
(v_c/220\kms)^2 h^{-1}{\rm kpc}$; the normalization is chosen to give
the right range of scalelengths observed (van der Kruit 1987), for $100
< v_c < 250 \kms$. We then use equation (4), integrated over $\nhi$, to
find the total number of systems per unit redshift above a given column
density. We take $\nhi = 10^{20.3} \cm^{-2}$ for the minimum column
density required to identify an absorption line as a damped system (see
Wolfe 1993).

  From the results shown in Figure 3, we see that the predictions of the
CDM model with $b=1$, 2, and the CDM$\Lambda$ model with $b=1$ may be
consistent with observations, if other effects do not reduce the cross
section significantly. In the CDM$\Lambda$ model, $n(z)$ at high $z$
depends strongly on the bias parameter $b$. If $b\approx 1.5$ (see Cen,
Gnedin \& Ostriker, 1993), the CDM$\Lambda$ model may have difficulty in
explaining the observed $n(z)$. But for the MDM model with $b\ge 1$,
there is a much clearer disagreement with observations. The discrepancy
at high redshift in this model is larger in $n(z)$ than in
$\Omega_{\rm D}(z)$ (see Fig. 2), because the average HI column density
predicted by the model is larger than observed. This is primarily due to
the fact that the central density of disks goes as $v_c^{-1}$ in our
disc model and that most mass is in small haloes at high redshifts (see
Fig. 2).

  \sect {4.~ DISCUSSION}

  In this paper, we have seen that the large amount of mass in neutral
gas observed in DL systems at high redshifts rules out the MDM model.
The reason is that the fraction of gas in collapsed halos which can
produce DL systems is too small. Even in the CDM and CDM$\Lambda$
models, one needs to assume that a large fraction of the gaseous
mass in galactic halos at $z\simeq 3$ is in disks of neutral gas.
Our calculated number of DL systems is more uncertain, since it
depends on the radial distribution of the gas in the disks. The
number of DL systems could be larger if the radius where the column
density is equal to $10^{20.3}\cm^{-2}$ (the threshold for DL systems)
was increased. This could be done while keeping most of the gaseous
mass close to the center,  where it would produce higher column density
systems, and would turn into the stellar disks in spiral galaxies at
present. The lower column density regions of the disks at large radius
might be destroyed before forming many stars, due to supernova
explosions, or infalling material resulting in shock-heating of the gas.
In fact, most HI disks in present galaxies are observed to flare at
large radius (van Gorkom 1993 and references therein).

  A second possibility is that there is a population of low column
density disks, which extend to large radius. The star formation rate
in such disks might be much lower than in regular spiral galaxies, and
they might either be destroyed when the galactic halos where they
reside merge, or might evolve into low surface brightness galaxies at
present, which could be difficult to detect (McGaugh 1994).

  Further constraints on the nature of DL systems may come through a
better determination of the column density distribution (which depends
on the surface density profiles and the maximum column density of the
disks), and from measurements of their transverse size, either in
gravitationally lensed quasars, or by direct imaging of the absorbing
galaxies (Briggs \etal 1989; Wolfe \etal 1992).

\bigskip
We would like to acknowledge Stacy McGaugh, T. Padmanabhan, Martin Rees
and Simon White for helpful discussions. HJM acknowledges support from a
SERC Postdoctoral Fellowship. JM thanks SERC for support in Cambridge,
and the W. M. Keck Foundation for support in Princeton.

\vfill\eject

\parskip .0cm plus .03cm

\sect{REFERENCES}

\mnras{Barcons, X., \& Fabian, A.~C. 1987}{224}{675}
\apj{Bardeen, J.~M., Bond, J.~R., Kaiser, N. \& Szalay A.~S. 1986}{304}{15}
\refbook{Bond, J.~R., 1993, in {\it The Environment and Evolution of
Galaxies}, eds. J.~M. Shull \& H.~A. Thronson, Kluwer: Dordrecht, p3}
\apj{Briggs, F.~H., Wolfe, A.~M., Liszt, H., Davis, M.~M., \& Turner,
K.~L. 1989}{341}{650}
\apj{Cen, R.~Y., Gnedin, N.~Y. \& Ostriker, J.~P. 1993}{417}{387}
\mnras{Efstathiou, G., 1992}{256}{43P}
\apj{Freeman, K.~C. 1970}{160}{811}
\aa{Jing, Y.~P., Mo, H.~J., B\"orner, G., \& Fang, L.~Z. 1994}{in press}
\apj{Klypin, A., Holtzman, J., Primack, J., \& Reg\"os, E. 1993}{416}{1}
\apj{Lanzetta, K.~M., Turnshek, D.~A., \& Wolfe A.~M., 1993}{in press}
\apjsup{Lanzetta, K.~M., Wolfe, A.~M., Turnshek, D.~A., Lu, L., McMahon,
R.~G., \& Hazard C., 1991}{77}{1}
\nature{McGaugh, S.~S., 1994}{in press}
\apj{Press, W.~H., Schechter, P. 1974}{187}{425}
\apj{Rao, S., \& Briggs, F.~H. 1993}{419}{515}
\apjsup{Sargent, W.~L.~W., Young, P.~J., Boksenberg, A., \& Tytler, D.
1980}{42}{41}
\refbook{Subramanian, K. \& Padmanabhan, T. 1994, submitted to Nature}
\apj{Tytler, D. 1987}{321}{49}
\aa{van der Kruit, P. C. 1987}{173}{59}
\refbook{van Gorkom J.H., 1993, in {\it The Environment and Evolution of
Galaxies}, eds. J.~M. Shull, H.~A. Thronson (Kluwer: Dordrecht), p345}
\apj{Walker, T.~P., Steigman, G., Schramm, D.~N., Olive, K.~A., \& Kang,
H.-S. 1991}{376}{51}
\mnras{White, S.~D.~M., Efstathiou, G., \& Frenk, C.~S. 1993}{262}{1023}
\mnras{White, S.~D.~M., \& Rees M.J. 1978}{183}{341}
\refbook{Wolfe A., 1993, in {\it Relativistic Astrophysics and Particle
Cosmology}, eds. C. W. Akerlof, M. A. Srednicki (New York Acad. of
Sci.: New York), p 281}
\apj{Wolfe, A.~M., Turnshek, D.~A., Lanzetta, K.~M., \& Oke, J.~B.
1992}{385}{151}
\vfill\eject

\parskip .5cm plus .1cm

\sect{ FIGURE CAPTIONS}

  {\bf Fig. 1:} The HI column density distribution function for damped
Lyman alpha systems. The five observational data points at high $\nhi$
are adopted from Lanzetta \etal (1991); the other one is adopted from
Tytler (1987). The curves (discussed in subsection 3.1) show the
predictions of a model in which all damped Lyman alpha absorbers are
exponential disks with central column densities $N_0$ shown in the
figure. The amplitudes of the curves are adjusted to the observation.

  {\bf Fig. 2:} The density of baryons
(scaled to zero redshift,in unit of present critical density)
in virialized halos as
a function of redshift, predicted by CDM (left panel), CDM$\Lambda$
(middle) and MDM (right). For each model, results for two bias
parameters are shown. For each bias parameter, results are shown for
three different intervals of halo circular velocities: $v_c> 50$, $50
-250$, $100-250\kms$, with the higher curve corresponding to the case
with a larger interval. The data points for the DL systems (solid
circles) are adopted from Lanzetta \etal (1993), adjusted according to
the cosmological model in consideration. The upper cross shows the
density parameter of baryons in stars of present-day galaxies. The lower
cross shows the HI mass in current spirals given by Rao \& Briggs (1993).

  {\bf Fig. 3:} The rate of incidence, $n(z)$, of damped Lyman alpha
absorptions predicted by CDM (left panel), CDM$\Lambda$ (middle) and
MDM (right). The gas mass in halos is assumed to settle in exponential
disks with scalelengths given by the Tully-Fisher relation (see text).

\vfill
\eject
\bye